# Increasing signal amplitude in electrical impedance tomography of neural activity using a parallel resistor inductor capacitor (RLC) circuit

J. Hope[1,2], Z. Aqrawe[3], M. Lim[1], F. Vanholsbeeck[2,4], A. McDaid[1]

[1]The Department of Mechanical Engineering, The University of Auckland, New Zealand
[2]Dodd Walls Centre for Photonic and Quantum Technologies, New Zealand
[3]The School of Pharmacy, The University of Auckland, New Zealand
[4]The Department of Physics, The University of Auckland, New Zealand

**Abstract**
**Objective:** To increase the impedance signal amplitude produced during neural activity using a novel approach of implementing a parallel resistor inductor capacitor (RLC) circuit across the current source used in electrical impedance tomography (EIT) of peripheral nerve.
**Approach:** Experiments were performed *in vitro* on sciatic nerve of Sprague-Dawley rats. Temporal changes to the impedance signal were used to correct data collected on the frequency response of the impedance signal in the range 4 – 18 kHz, then a frequency range with significant capacitive charge transfer was selected for experiment with the RLC circuit. Design of the RLC circuit was performed in electrical circuit modelling software, aided by *in vitro* impedance measurements on nerve and nerve cuff in the range 5 Hz to 50 kHz.
**Main results:** Temporal changes to the impedance signal amplitude and the delay between stimulation and impedance signal were significant, with the former decreasing by 44 % and the latter increasing by 28 to 50 % over 16 minutes 40 seconds. The frequency response of the impedance signal across 4 – 18 kHz showed maximum amplitude at 6 to 8 kHz, and steady decline in amplitude between 8 and 18 kHz with -6 dB reduction at 14 kHz. The frequency range 17 +/- 1 kHz was selected for the RLC experiment. The RLC experiment was performed on four subjects using an RLC circuit designed to produce a resonant frequency of 17 kHz with a bandwidth of 3.6 kHz, and containing a 22 mH inductive element and a 3.45 nF capacitive element with +0.8/-3.45 nF manual tuning range. With the RLC circuit connected, relative increases in the impedance signal (+/-3σ noise) of 44 % (+/-15 %), 33 % (+/-30 %), 37 % (+/-8.6 %), and 16 % (+/-19 %) were produced.
**Significance:** The increase in impedance signal amplitude at high frequencies, generated by the novel implementation of a parallel RLC circuit across the drive current, improves spatial resolution by increasing the number of parallel drive currents which can be implemented in a frequency division multiplexed (FDM) EIT system, and aids the long term goal of a real-time FDM EIT system by reducing the need for ensemble averaging.

## 1. Introduction

Electrical Impedance Tomography (EIT) is an imaging modality which probes the sample using several drive currents, each applied at a different orientation with respect to the sample, while recording the resultant boundary voltages. Through Calderon's inverse problem, the boundary voltages are used to reconstruct estimates of the distribution of electrical impedance through the sample. A higher number of drive currents increases the number of measurements, and, in turn, the spatial resolution of the resolved impedance distribution [1]. EIT imaging of neural activity is made possible by the change in the membrane resistance during generation and propagation of action potentials in the nervous system, which modifies the tissue impedance [2-5]. Researchers have used this principle to image neural activity in the brain using epicortical electrode arrays [6-9] and in the peripheral nerve using nerve cuff electrode arrays [10, 11].

In EIT, the requirement to obtain information using multiple drive currents necessitates some form of multiplexing so that the boundary voltage information unique to each drive current orientation can be resolved. One commonly employed method is time division multiplexing (TDM) [6-8, 10, 11], where a single drive current is applied through each drive electrode pair sequentially in time. While TDM is suitable for studies where repeated neural activity, or stimulus, is available, it is inadequate for studies where the target neural activity must recorded in real-time. Frequency division multiplexing (FDM) of drive currents [12], on the other hand, is theoretically compatible with real-time applications, and has been demonstrated using two multiplexed drive currents in initial studies on epileptiform activity in rat brain [13] and fast neural activity in peripheral nerve [14]. The drawbacks of FDM are the lower signal to noise ratio (SNR) due to each drive current being a noise source and the reduced drive current amplitudes required to avoid stimulating neural tissue, and the limited number of drive currents which can be implemented due to the frequency response roll-off in impedance of neural tissue [10, 15-17].

The origin of the frequency response roll-off in impedance is capacitive charge transfer across the neuron membrane or myelin sheath, depending on the neural tissue under study, and is of concern to EIT as it reduces the amount of resistive current passing through ion channels to which EIT is sensitive [4, 5]. The different cytology of brain and nerve tissues produce different electrical impedance and frequency response roll-off characteristics. Brain tissue exhibits the largest transient change in impedance during neural activity, which we may term the 'impedance signal', between 1 and 2 kHz and minimal signal above 3 kHz [15, 16], whereas peripheral nerve tissue with a transverse current, across the nerve fibres, shows the largest impedance signal between 6 and 10 kHz [10], though frequencies above 15 kHz have not yet been characterised in experiments. The impedance signal in a peripheral nerve using a longitudinal current, parallel to nerve fibres, has not been characterised, though estimates from the frequency response roll-off of inactive tissue indicate 20 kHz may be the upper limit [17]. In all neural tissues, the frequency spacing between two FDM drive currents is determined by the frequency components of the modulated neural activity [14], with faster activity containing higher frequency components. When considered together, such as in an FDM EIT system, the frequency spacing requirement and frequency response roll-off impose a limit on the number of drive currents, and so restricts the achievable spatial resolution in real-time EIT.

In the current study, we develop and test a parallel RLC circuit on an EIT current source using a multi-channel nerve cuff implanted on sciatic nerve of rat *in-vitro* in an attempt to develop FDM EIT technology towards real time imaging. In a parallel Resistor-Inductor-Capacitor (RLC) circuit, at the frequency where the impedance of the capacitive and inductive branches are equal, the effective current is determined solely by the resistive branches. In the case of EIT on neural tissue, this resistive branch includes the resistive current through the ion channels to which EIT is sensitive. The frequency response roll-off of the impedance signal during evoked neural activity is characterised with longitudinal current to identify a target frequency range where significant capacitive charge transfer occurs. Estimates of the resistive and capacitive components of peripheral nerve are acquired from software fitting of *ex-vivo* impedance magnitude and phase measurements to equivalent electrical circuits. A RLC circuit is then designed, based on the equivalent electrical circuit and target frequency range, and used to increase the impedance signal during evoked neural activity *in-vitro* by counteracting the capacitive charge transfer across the myelin sheath. This novel implementation of a parallel RLC circuit across the drive current increases the impedance signal at high frequencies, thereby improving spatial resolution by increasing the number of parallel drive currents which can be implemented in an FDM EIT system, and aiding the long term goal of a real-time FDM EIT system by reducing the number of epochs in ensemble averaging. The benefits of the parallel RLC circuit implementation in FDM EIT extend beyond neural tissue to any biological tissue which exhibits frequency response roll off.

## 2. Methods

*2.1 Tissue preparation and handling*

Animal procedures were approved by the University of Auckland Animal Ethics Advisory Committee. A total of nine subjects were used in the current study, all Sprague Dawley rats, in the weight range 450 – 650 g, male, and were provided as part of routine population management. All subjects were used in *in-vitro* experiments after being euthanized using carbon dioxide followed by cervical dislocation.

To access the sciatic nerve, subjects were placed in the supine position, the left hind leg was de-gloved of skin, and an incision was made down the posterior side of the leg. Tissue of the Gracilis muscle was cut away to aid implantation of the nerve cuff around the sciatic nerve. The nerve cuff was implanted by lifting the sciatic nerve away from the adjacent muscle tissue, using curved tip tweezers and without grasping the nerve to avoid causing damage, then sliding the nerve cuff under and around the nerve. Exposed tissue was perfused with a 0.01 M phosphate buffered solution which was oxygenated with Carbogen 5 gas and heated to 38 ºC. The left hind paw was fixed in place using a strap around the hock to minimise movement artefacts, though previous experiments indicated that movement induced by paw stimulation was minimal [14].

*2.2 Nerve cuff*

The nerve cuff electrode array was produced by *The EIT Research Group, The Department of Medical Physics and Biomedical Engineering, University College London,* using the fabrication method described in [11]. The electrode array contained 2 rings of 14 stainless steel electrodes, with each electrode 1.1 x 0.11 mm in area and spaced at a 26 ° pitch, and the two rings spaced 2 mm apart centre to centre. This centre to centre distance was extended to 4 mm (an edge to edge spacing of 3mm) prior to assembly of the nerve cuff using method described in [14]. Electrodes were cleaned using 35% Hydrochloric acid, then coated with poly(3,4-ethylenedioxythiophene):p-toluene sulfonate (PEDOT-pTS), using the three electrode galvanostatic electrochemical deposition method described in [11] (Biologic Science Instruments® VSP-300), to reduce the contact impedance from 20+/-5.6 to 800 +/- 100 Ω. The nerve cuff was assembled by affixing the array using high strength silicon glue to the inside surface of a 1.2 x 4 x 9 mm silicon tube, and containing an opening extending the length of one side to facilitate implantation around the nerve [14]. The nerve cuff was soaked in PBS for one hour prior to being implanted.

      Three nerve cuffs were used across all experiments. Prior to each experiment, three point impedance magnitude measurements were acquired with a 4 µA current controlled sinusoidal signal at 6 kHz for each of the 28 electrodes to monitor electrode condition. Highly fouled or damaged electrodes, as determined by an excessive voltage magnitude, were not used in experiments. After each experiment, the electrode array was cleaned with isopropyl alcohol and then stored in air.

      The nerve cuff was divided into 14 pairs of electrodes, each comprising one electrode on the two electrode rings located at the same angular position. This configuration is termed 'Longitudinal 0°/0°' in [18], where Longitudinal indicates the current direction – parallel to nerve fibres – and the two angular references indicate the angular offset between drive and measurement electrode pairs, respectively, and was selected to optimise the signal to error ratio based on modelling in [18].

*2.3 Experiment apparatus*

A total of four boundary voltages, or two differential measurement pairs, were acquired. In all experiment results, data from the electrode pair with the highest SNR is presented. Boundary voltage measurements were filtered using an anti-aliasing filter, comprised of two in-series 2nd order low-pass filters of Sallen-Key topology, producing a 25 kHz cut-off frequency (-3 dB) and Q = 0.69. Filtered signals were logged in differential configuration on two data acquisition modules (National Instruments CompactRIO® NI9205) at 167 kS/s/ch sample rate. Sample rate was selected to be at least a factor of 9 times that of the highest drive current frequency present.

      Two stainless steel pins, used to administer stimulation pulses, were placed between toes 1-2 and 4-5 of the left hind paw [14], and a third stainless steel pin, which provided a ground path for the data acquisition and filtering hardware via a 460 kΩ resistor, was placed through the contralateral hind paw. An isolated pulse stimulator (A-M Systems Model 2100) administered anode-leading, biphasic stimulation pulses, with 200 µs pulse width per phase and +/-5 mA amplitude, when digitally triggered every 0.5 s [10, 14] (National Instruments CompactRIO NI9401). Triggering and recording were synchronised through the LABVIEW FPGA interface, with data recorded for 50 ms windows every 500 ms, and the stimulation pulse triggered at 25 ms through each recording window. Recorded data were saved directly to .txt files for signal processing after completion of experiments.

      With the exception of the experiment using the RLC circuit, EIT drive currents were generated using a custom PCB with 6 parallel circuits each containing a waveform generator and Howland circuit current controller, developed by *The EIT Research Group at The Department of Medical Physics and Biomedical Engineering, University College London,* and available for download from https://github.com/EIT-team. A different frequency was programmed for each of the six channels prior to experiments, using an Arduino Micro, to allow manual switching of drive current frequency during experiments.

      In RLC circuit experiments a custom current controller was produced, based on the component layout of the *University College London EIT Research Group* current controller design, using LM358A operational amplifiers (Texas Instruments). The input signal was provided to the current controller by a benchtop waveform generator (Thurley Thandar Instruments TG2000) to allow manual tuning of frequency. The RLC circuit was

connected in parallel across the two current source outputs, with a double-pole-double-throw (DPDT) switch implemented to allow manual switching between connection and disconnection of the RLC circuit, Fig 1.

Phase randomisation, to remove current source artefacts [19], was achieved in the custom PCB implemented waveform generator through an inbuilt inaccuracy of less than 1 Hz, and in the benchtop waveform generator by setting the frequency output to end in 0.2 Hz.

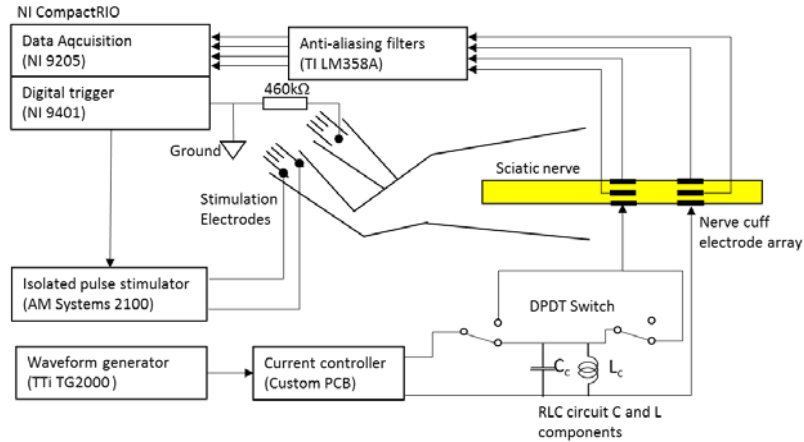

Figure 1: Schematic of the apparatus used in RLC circuit experiments (experiment 5). In all other experiments the waveform generator was implemented on board the current controller PCB (https://github.com/EIT-team) containing six parallel channels with programmable output frequencies.

*2.4 Data processing*

In all experiments, data were recorded in 50 ms windows, termed an 'epoch', centred on a stimulation pulse, and spaced 0.5 seconds apart based on the stimulation rate. Impedance signals were extracted from each epoch using the method described in [14], which, in order, involves: removal of DC offset; application of a Tukey window with 40 % tapered section; $6^{th}$ order, zero phase shift, 1.5 kHz bandwidth, Band Pass Filter (BPF); Hilbert Transform; and, conversion to a percentage change in impedance $z_\%(t)$ with respect to the inactive nerve state using equation 1:

$$z_\%(t) = 100 \left( \frac{V(t) - V_i}{V_i} \right) \qquad (1)$$

where the boundary voltage in the inactive nerve state, $V_i$, was calculated as the average across the time spanning 10 to 5 ms prior to the stimulus pulse.

All epochs were processed with the above steps, then several hundred epochs of data were grouped together into data sets, based on the sampling method employed, and ensemble averaged to improve the SNR according to standard practice in impedance signal recordings [10, 14]. An example of this collection and ensemble averaging methodology is shown in Fig 3, using, as an example, four recordings of action potentials in response to a stimulation pulse. Grouping of epochs into sets, the 'sampling method', varied between experiments, Fig 4, and is explained in the subsequent sections.

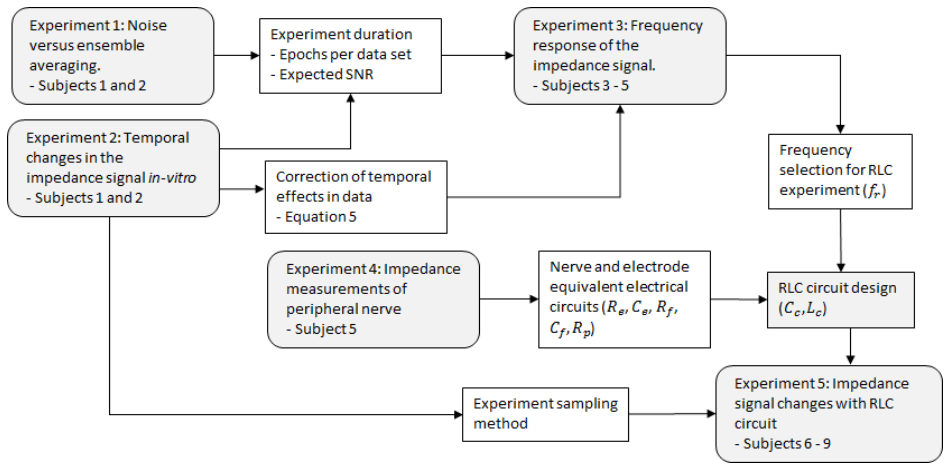

Figure 2: Layout of the five experiments and RLC circuit design performed as part of the current study, showing how results from each experiment informed the others.

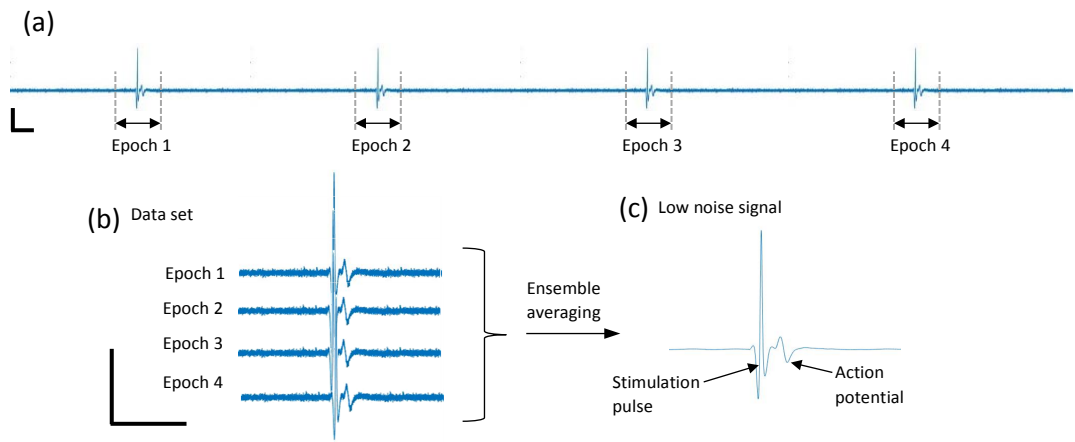

Figure 3: Example of ensemble averaging of data wherein four epochs, originally spaced 0.25 seconds apart (a), are grouped into a data set (b) and then ensemble averaged together to produce a single signal with reduced noise (c). Horizontal and vertical bars represent 25 ms and 0.5 mV, respectively.

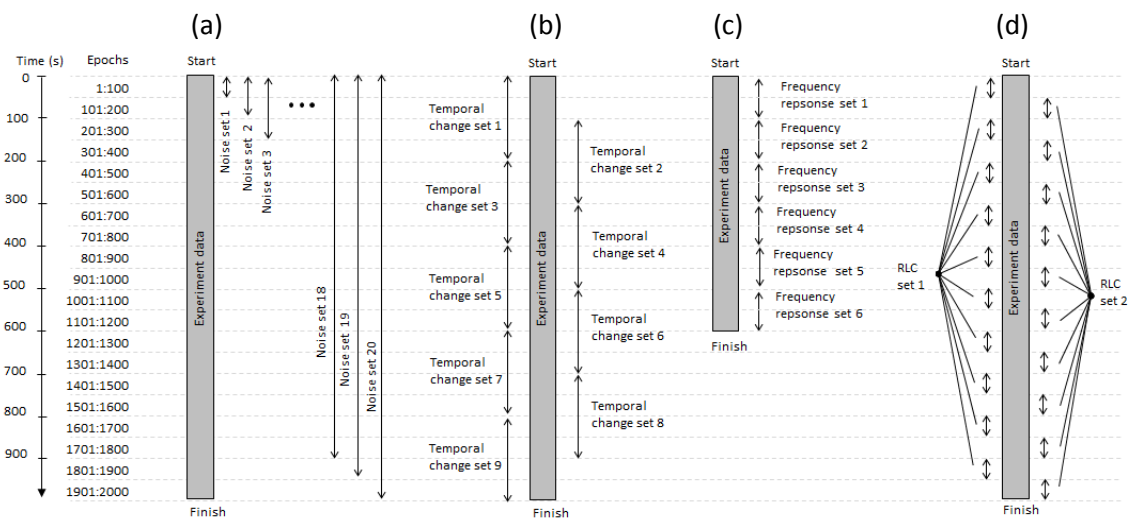

Figure 4: Grouping of experimental data into sets for characterisation of system noise versus ensemble averaging in experiment 1 (a); characterisation of the temporal changes in the impedance signal in-vitro signal in experiment 2 (b); characterising the frequency response of the impedance signal in experiment 3 (c); and to evaluate the peak amplitude of the impedance signal with and without the RLC circuit in experiment 5 (d). The time lapse between neighbouring epochs is 0.5 s.

*2.5 Experiment procedures*

A total of five experiments were performed across the nine animal subjects, Fig 2. In experiment 1, the noise after data processing and ensemble averaging was characterised with respect to the number of epochs within the data set. In experiment 2, temporal changes to the impedance signal were characterised. In experiment 3, the frequency response of the impedance signal was characterised, and then corrected for temporal changes in the impedance signal using results of experiment 2. The duration of experiment 3 was selected to provide sufficient SNR by balancing the size of data sets collected at each frequency, using results of experiment 1, with the temporal period which showed the maximum impedance signal amplitude, using results of experiment 2. Results of experiment 3 were used to identify frequencies considered comfortably within the frequency response roll-off zone. In experiment 4, impedance measurements were acquired on a sciatic nerve within a nerve cuff, and of a nerve cuff in PBS, and then these were used to estimate values of the capacitive and resistive components in equivalent electrical circuits.

The equivalent electrical circuits and the target frequency from experiments 4 and 3, respectively, were used to select the capacitive and inductive components for the design of the RLC circuit. In experiment 5, changes in the impedance signal were characterised by comparing data with RLC circuit on and RLC circuit off. The sampling method was selected to mitigate errors produced by temporal effects, and these errors were estimated, using data from experiment 2.

2.5.1 Experiment 1: Noise versus ensemble averaging

Application of a repeated, time-locked stimulus allowed ensemble averaging of multiple epochs to be used during data processing to reduce noise and improve the SNR. The relationship between signal noise and the number of ensemble averaged epochs was characterised *in-vitro* on two animals (subjects 1 and 2) using a 16 µA amplitude [14], 6 kHz drive current applied for 2000 epochs. Each epoch was processed using the method described in *section 2.4*, then data were ensemble averaged in sets ranging in size from 100 to 2000 epochs in increments of 100, Fig 4a. The standard deviation of noise was calculated for each ensemble averaged set across the time window 15 to 23 ms which, due to being prior to stimulus pulse at 25 ms, was expected to contain no neural activity.

2.5.2 Experiment 2: Temporal changes in the impedance signal *in-vitro*

With *in-vitro* experiments on cadavers, the tissue health is expected to deteriorate with time, following euthanasia of the animal, due to the lack of natural perfusion and cooling of the tissue. Therefore, we characterized temporal changes in the impedance signal on two animals (subjects 1 and 2) across 2000 epochs, or 16 minutes 40 seconds, using the same raw data as that obtained for experiment 1 (Noise versus ensemble averaging experiments, *section 2.5.1*.). Each epoch was processed using the method described in *section 2.4*, then separated into sets of 400 epochs (200 s), each overlapping the previous set by 200 epochs (100 s) Fig 4b, and ensemble averaged. The temporal change in the impedance signal was fitted with a 2$^{nd}$ order polynomial, which was normalised against the maximum value.

2.5.3 Experiment 3: Frequency response of the impedance signal

The frequency dependence of the impedance signal during neural activity with a longitudinal EIT current was measured on three animals (subjects 3 – 5). For each subject, a 16 µA amplitude drive current was applied sequentially at six frequencies between 4 and 18 kHz. Each frequency was applied for 200 epochs, with the order of the frequencies applied randomized, Fig 4c. Data were processed using the method described in *section 2.4*, then corrected for temporal changes using the 2$^{nd}$ order polynomial fit obtained from experiment 2, *section 2.5.2*.

Results of the impedance signal frequency response experiments were used to select a drive current frequency which was comfortably within the frequency response roll-off, and so expected to contain significant capacitive current across the myelin sheath. The selected drive frequency range was used to design the RLC circuit.

2.5.4 Experiment 4: Impedance measurements of peripheral nerve

Estimates of the resistive and capacitive components in equivalent electrical circuits of peripheral nerve and nerve cuff electrodes were obtained to aid design of the RLC circuit. A simplified equivalent electrical circuit of a nerve contains a capacitor, $C_f$, and resistor, $R_f$, in parallel to model the nerve fibres, and a resistor, $R_p$, in series to model the passive tissues [17, 20], Fig 5a. The equivalent electrical circuit of an electrode in solution or in contact with biological tissue, and at frequencies where contributions from the Warburg impedance due to ion diffusion is negligible, contains a capacitor, $C_e$, and resistor, $R_e$, in parallel [21], Fig 5b. To avoid contributions from the Warburg impedance, only frequencies greater than 2 kHz were used in fitting the electrical circuits to impedance data.

At the conclusion of one *in-vitro* experiment, the nerve, still within the nerve cuff, was explanted and transferred to PBS solution. Impedance measurements were acquired with a bench-top impedance analyser (Biologic Science Instruments® VSP-300) between the current injection electrode pair, across the frequency range 5 Hz to 50 kHz, first with the nerve in the nerve cuff and in solution, and then with the only nerve cuff in the PBS solution. Impedance data from the latter was subtracted from the former to produce an estimate of the impedance of the nerve. The estimate of the impedance of the nerve alone, and the impedance data of the nerve in nerve cuff, were fitted to equivalent electrical circuits, Fig 5a and 5b respectively, using software (Biologic Science Instruments® EC-Lab V11.25). In this software, the user defines the circuit and components and the software generates values for each component.

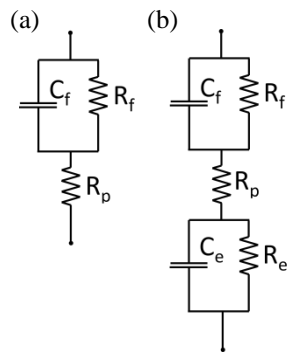

Figure 5: Equivalent electrical circuit for peripheral nerve (a) and peripheral nerve in a nerve cuff (b), showing a capacitor, $C_f$, and resistor, $R_f$, in parallel to model the nerve fibres, a resistor, $R_p$, to model the passive nerve tissues, and a capacitor, $C_e$, and resistor, $R_e$, in parallel to model the electrode at frequencies with negligible Warburg impedance.

2.5.5 RLC circuit design

To achieve maximum sensitivity to a change in tissue resistance during neural activity, the resistive component, $R$, in the RLC circuit should predominantly be the resistive element of the neural tissue. Therefore, no additional resistors were added during RLC circuit design. The remaining two components, the capacitive, $C$, and inductive, $L$, components, were selected to produce the desired resonant frequency, $f_r$, using equation 2 and the target frequency range identified in experiment 3 (Frequency response of the impedance signal, *section 2.5.3*):

$$f_r = \frac{1}{2\pi\sqrt{LC}} \qquad (2)$$

The bandwidth of RLC circuit response was limited to values above 1 kHz to allow a frequency sweep to be performed in 100 Hz increments during RLC experiments. A higher Quality Factor, equation 3, produces a narrower bandwidth, $BW$, of the pass band; where the bandwidth is defined as the difference between the upper and lower frequencies at the -3 dB amplitude reduction from the amplitude at the resonant frequency, and can be calculated from $f_r$ and $Q$, equation 4:

$$Q = R\sqrt{\frac{C}{L}} \qquad (3)$$

$$BW = \frac{f_r}{Q} \qquad (4)$$

The capacitive component, $C$, was assumed to contain contributions from the neural tissue capacitance, $C_f$, electrode capacitance, $C_e$, and any physical capacitive component added to the RLC circuit, $C_c$; whereas the inductive component, $L$, was assumed to contain only a contribution from the physical inductive component added to the RLC circuit, $L_c$. Values of the inductance, $L_c$, and capacitance, $C_c$, components were selected by modelling circuits (Mathworks® MATLAB 2018b, LTSpiceIV 4.23b), Fig 6, using the component values identified from impedance measurements in experiment 4 (Impedance measurements of peripheral nerve, *section 2.5.4*).

Three RLC circuits with different bandwidths were designed and built. In each RLC circuit the capacitances were implemented as six capacitors in parallel, with values of approximately 65, 30, 15, 15, 5, and 3 % of the target capacitance, $C_c$, to allow manual tuning within the range +33/-100 % using jumpers in the event of the neural tissue capacitance varying between subjects. Inductor values were fixed in each RLC circuit, and selected based on commonly available component values (Digikey®).

The impedance signal frequency response of the equivalent electrical circuits of the nerve in nerve cuff, Fig 5b, and RLC circuit, Fig 6, were modelled (LTSpiceIV 4.23b) across the frequency range 1 to 100 kHz. The inactive nerve condition, $V_i$, used the fibre resistance value, $R_f$, obtained from impedance measurements, *section 2.5.4*, whereas for the active nerve condition, $V_a$, the fibre resistance value was reduced by 0.1 % [14]. The impedance signal was then calculated as $z_\% = 100(V_i - V_a)/V_a$ to provide an estimates of the expected change in impedance signal when using the designed RLC circuits.

2.5.6 Experiment 5: Impedance signal changes with RLC circuit

Impedance signal changes with the RLC circuit were characterised on four animals (subjects 6 – 9). After the nerve cuff was implanted, drive current frequencies were swept manually from 10 to 30 kHz with 100 Hz resolution and the peak impedance recorded. The capacitance of the RLC circuit was then adjusted upwards or downwards accordingly, using the manual jumpers, and drive current frequencies swept again until the capacitance settings produced a resonant frequency to within +/-1 kHz of the target frequency. The circuit tuning process took less than 1 minute. Once the circuit settings were identified, data were collected with and without the RLC circuit connected to the drive current source, alternating in sets of 100 epochs each for a total of 2000 epochs, Fig 4d. Switching between conditions with and without the RLC circuit was achieved by manually toggling the DPDT switch connecting the RLC circuit to the current controller, Fig 1.

The error produced from this sampling method, of compiling two data sets by switching every 100 epochs, was estimated by applying the sampling method to the data sets from subjects 1 and 2 in experiment 2 (Temporal changes in the impedance signal in-vitro, *section 2.5.2*) and comparing the resultant impedance signals. Furthermore, half of subjects commenced with the RLC circuit on, and the other half with the RLC circuit off to mitigate errors from the sampling method.

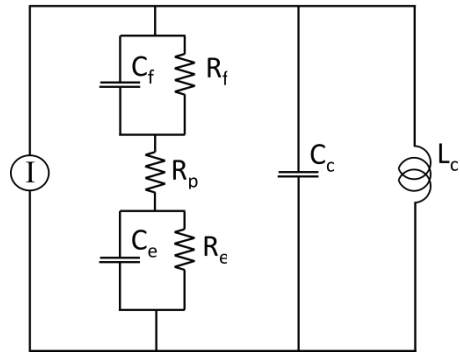

Figure 6: RLC circuit, with current source, $I$, containing the peripheral nerve in a nerve cuff from Fig 5b, as well as the inductance, $L_c$, and capacitance, $C_c$, components used to tune the RLC circuit in experiment 5.

## 3. Results

*3.1 Experiment 1: Noise versus ensemble averaging*

The noise in boundary voltage signals reduced with increasing number of ensemble averaged epochs, following an approximately linear relationship when plotted on log-log scale, Fig 7. The largest increase in SNR was, therefore, generated by ensemble averaging of the first few epochs, with a decreasing effect as more epochs were added to an ensemble averaged set. A typical impedance signal amplitude of 0.02 to 0.08 % and boundary voltage amplitudes of approximately +/- 150 mV, from a study with the same nerve cuff and electrode configuration [14], would produce a transient change in boundary voltage with 30 to 120 µV peak amplitude. Using peak to peak noise values, a target SNR of 3 would require at least 2000 and 150 epochs, respectively, for 30 and 120 µV amplitude signals.

The selection of 200 and 400 epochs per set for frequency response and temporal changes experiments, respectively, are expected to produce SNR > 3 for large impedance signals but SNR < 3 for smaller impedance signals. The reduced boundary voltage amplitude observed at higher frequencies in [17] may render even a set size of 1000 epochs, used in RLC experiment, incapable of producing an SNR > 3.

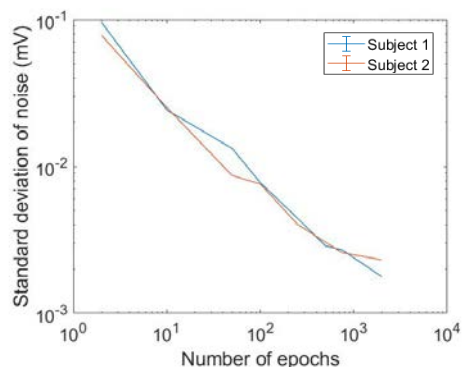

Figure 7: The system noise on the boundary voltage measurements reduces with an increasing number ensemble averaged epochs, following an approximately linear relationship when plotted on a log-log scale.

*3.2 Experiment 2: Temporal changes in the impedance signal in-vitro*

In subject 1, Fig 8a, the peak amplitude of the impedance signal increased from -0.071 to -0.081 % between the first and third sets (100 and 300 s) and then decreased thereafter to a minimum of -0.046 %, a relative decrease of 43 %, Fig 8c. In subject 2, Fig 8b, the impedance signal showed a small increase from 0.091 to 0.093 % between the first and second sets (100 and 200 s) and then decreased through the course of the experiment to a minimum of 0.052 % at 900 seconds, a relative decrease of 44 %, Fig 8c. With the exception of the first data point, at 100 s from experiment start, the temporal change in impedance signal amplitude across subjects 1 and 2 showed reasonable agreement. A second order polynomial fitted to both data sets, equation 5:

$$z_\% = -7.12 \times 10^{-8} \, t^2 + 2.91 \times 10^{-5} t + 0.0809 \qquad (5)$$

where $z_\%$ is the impedance signal amplitude and $t$ the time lapsed since start of the experiment, showed minimal change in impedance signal amplitude from 100 to 400 s, and then an increasing rate of decline between 400 and 900 s, Fig 8c.

In subjects 1 and 2, the delay between the start of the stimulus and the peak amplitude of the impedance signal increased linearly with time since the experiment start, Fig 8d, although at different rates. The total increase in delay observed across the course of subjects 1 and 2 was 28 and 50 %, respectively.

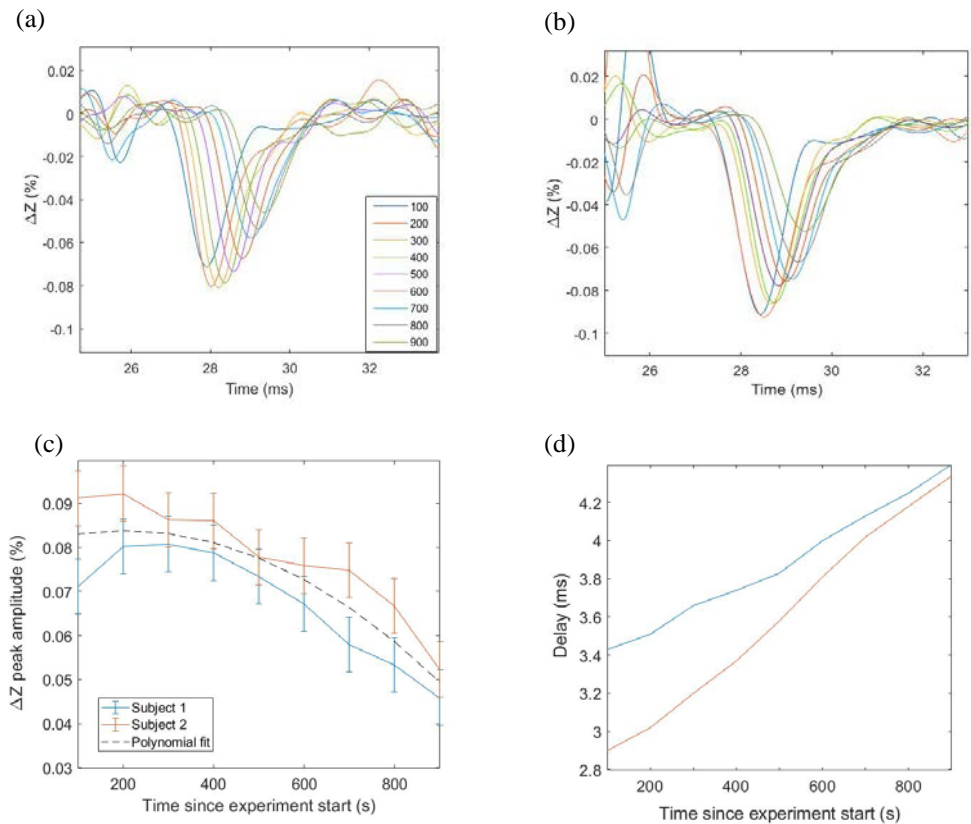

Figure 8: Impedance signals extracted from subjects 1 (a) and 2 (b) for each of the nine sets of epochs which are spaced in 100 second increments from the start of the experiment. The peak amplitude of the impedance signal, and the delay between the stimulus pulse at 25 ms and the peak of the impedances signal, are both analysed against the time since the experiment started in (c) and (d), respectively, and show clear changes as time progresses from 100 to 900 seconds.

*3.3 Experiment 3: Frequency response of the impedance signal*

Impedance signal amplitudes from subject 3 were consistently higher than those in subjects 4 and 5.. After correction for temporal changes, the frequency response of all three subjects showed an increase in the peak amplitude of the impedance signal between 4 and 6 kHz, and decrease between 8 and 18 kHz, Fig 9. At 16 and 18 kHz, SNRs < 3 were observed, however, good agreement between the three subjects at these frequencies provided good confidence in the values despite the low SNR. The trend in impedance signal against frequency in all three subjects showed good agreement with one another across the entire 4 – 18 kHz frequency range.

A target frequency range of 17 +/- 1 kHz was selected for the RLC circuit experiment as this range was considered comfortably within the impedance frequency response roll-off, indicating significant current crossing the myelin sheath through capacitive charge transfer.

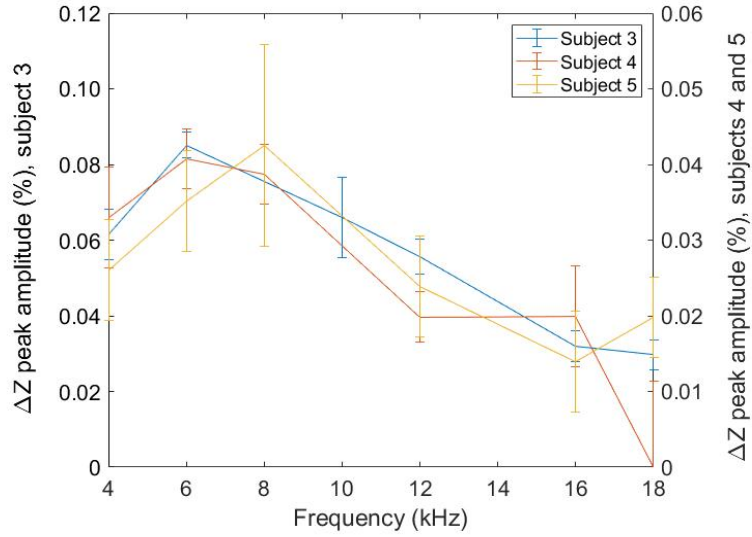

Figure 9: The frequency response of the impedance signal amplitude after correction for temporal change effects using the polynomial fit, equation 5. Error bars are the peak to peak noise level in the measurement and so indicate the measurement uncertainty. There is good agreement between all three experiments, and clear frequency response roll-off between 8 and 18 kHz.

*3.4 Experiment 4: Impedance measurements of peripheral nerve*

Nyquist impedance plots of the nerve cuff in PBS and of the nerve in nerve cuff both exhibited local minima at approximately 100 Hz and increasing real and imaginary parts at frequencies < 100 Hz, with the latter characteristic of contributions from the Warburg impedance due to ion diffusion, Fig 10. The frequency range of 2 to 50 kHz was considered sufficiently far from this minima to render these contributions negligible. Fitting of the equivalent electrical circuit, Fig 5b, to impedance data from the nerve in nerve cuff produced a fit accuracy with $\chi^2/|Z| = 5.413 \times 10^{-3}$ for the following component values: nerve fibre capacitance $C_f = 0.75\ nF$, and resistance $R_f = 8.7\ k\Omega$; passive nerve tissue resistance $R_p = 1.8\ k\Omega$; electrode capacitance $C_e = 80\ nF$, and resistance $R_e = 0.86\ k\Omega$. The total resistance in the RLC circuit, R, was, therefore, $R_f + R_p + R_e = 11.26\ k\Omega$. Fitting of equivalent electrical circuits, Fig 5a, to impedance data from the nerve estimate, obtained by subtracting the nerve cuff in PBS impedance data from the nerve in nerve cuff impedance data, produced a fit accuracy of $\chi^2/|Z| = 7.065 \times 10^{-3}$ for the following values: nerve fibre capacitance $C_f = 0.8\ nF$, and resistance $R_f = 7.9\ k\Omega$; passive nerve tissue resistance $R_p = 1.7\ k\Omega$, Table 1. The low $\chi^2/|Z|$ values for both circuits indicate good agreement between the data and equivalent electrical circuits.

*3.5 RLC circuit*

3.5.1 RLC circuit design

Modelling of RLC circuits, using equations 2 - 4 with nerve and electrode values listed in Table 1, generated a range of suitable inductance and capacitance values for the RLC circuit to achieve the 17 kHz target resonant frequency, Fig 11a. Limiting values to those which produce a bandwidth greater than 1 kHz limited inductance values to 6 mH or greater, Fig 11a. Three inductances, $L_c$: 13, 22, and 44 mH, selected based on commonly available component values (Digikey®), generated corresponding capacitance values, $C_c$, of 6.25, 3.45, and 1.45 nF, respectively. The resulting theoretical bandwidth of these three component combinations were 2.2, 3.6, and 7 kHz.

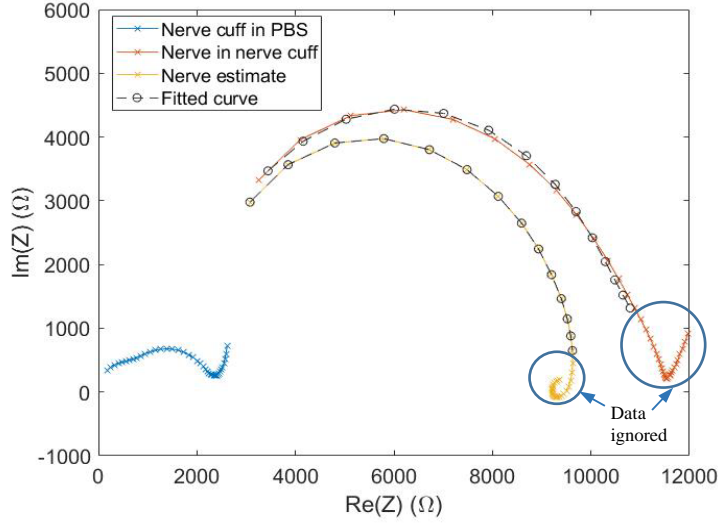

Figure 10: Nyquist impedance plots for the nerve cuff in PBS, and the nerve in nerve cuff, as well as the estimate of the Nyquist impedance of the nerve which was calculated from the latter minus the former. The curves fitted to the nerve in nerve cuff data and to the nerve estimate impedance, and used to generate component values for the equivalent electrical circuits in Fig 5a and b, shows good agreement between the selected frequency range 2 to 50 kHz. Data in the 5 Hz to 2 kHz range, encircled, was ignored when fitting circuits.

Table 1: Parameter description and values used in modelling

| Description | Parameter | Value |
|---|---|---|
| **Nerve tissue** | | |
| Nerve fibre capacitance | $C_f$ | 0.75 - 0.8 nF |
| Nerve fibre resistance | $R_f$ | 7.9 - 8.7 kΩ |
| Passive nerve tissue resistance | $R_p$ | 1.7 - 1.8 kΩ |
| **Nerve cuff electrodes** | | |
| Electrode capacitance | $C_e$ | 80 nF |
| Electrode resistance | $R_e$ | 0.8 kΩ |
| **RLC circuit configuration 1 ($BW = 2.2$ kHz)** | | |
| Capacitance component | $C_c$ | 6.25 nF |
| Inductance component | $L_c$ | 13 mH |
| **RLC circuit configuration 2 ($BW = 3.6$ kHz)** | | |
| Capacitance component | $C_c$ | 3.45 nF |
| Inductance component | $L_c$ | 22 mH |
| **RLC circuit configuration 3 ($BW = 7$ kHz)** | | |
| Capacitance component | $C_c$ | 1.45 nF |
| Inductance component | $L_c$ | 44 mH |

The modelled frequency response of the impedance signal in the nerve and nerve cuff circuit, Fig 5b, across the frequency 1 to 100 kHz, and with parameter values in Table 1, showed a local maximum of -0.08 % at 4 kHz, with a slight decrease between 4 – 2 kHz and a significant decrease, due to the frequency response roll-off, between 4 and 100 kHz. At 17 kHz, the impedance signal in the nerve and nerve cuff circuit was -0.066 %, whereas in all three RLC circuits it was -0.077 %; a relative improvement of 17 %, Fig 11b. The impedance signal with the RLC circuits was less than that of the maximum in the nerve and nerve cuff circuit, of -0.08 % at 4 kHz, due to the passive nerve tissue resistance, $R_p$, and electrode resistance, $R_e$. With these two component values set nominally low, to 1 Ω, the impedance signal in the RLC circuit was equal to the maximum value in the nerve and nerve cuff circuit, and constant at all frequencies.

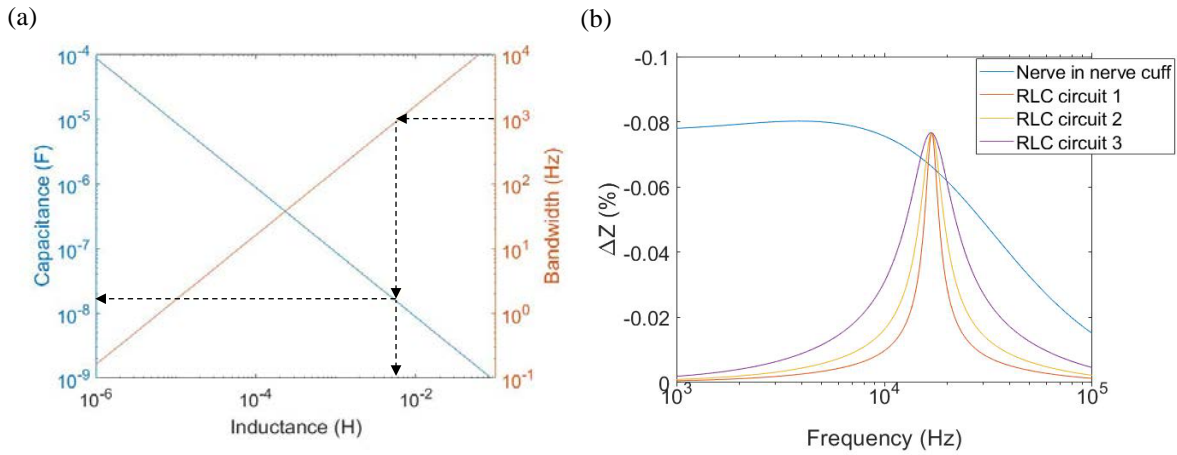

Figure 11: Log-log plot showing the relationship between the RLC capacitance component, $C_c$, RLC inductance component, $L_c$, and bandwidth (a) for the circuit shown in Fig 6 and values in Table 1 producing a 17 kHz resonant frequency. Only Bandwidths greater than 1 kHz were considered. The frequency response (b) of the impedance signal generated from a 0.1 % change in the nerve fibre resistance in equivalent electrical circuits of the nerve and nerve cuff circuit and the three RLC circuits with values shown in Table 1.

3.5.2 Estimate of error from sampling methodology

Application of the sampling method implemented in experiment 5 (Impedance signal changes with RLC circuit, *section 2.5.6*), Fig 4d, to data from subjects 1 and 2 produced impedance signal amplitudes of -0.30 and -0.33 %, and -0.55 and -0.66 %, in sets 1 and 2, respectively; a relative increase of 10 and 20 % in set 2 in subjects 1 and 2, respectively, Fig 12. Peak to peak noise in both experiments was +/- 0.005 %.

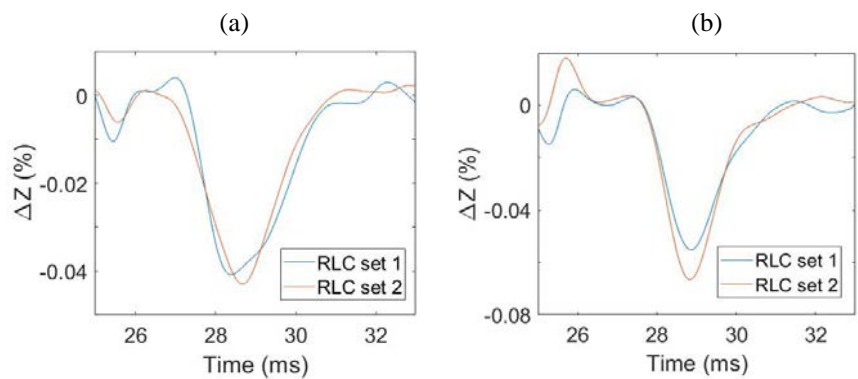

Figure 12: Averaging method employed in RLC circuit experiments, Fig 4d, where two sets of 1000 epochs each are interspliced at 100 epoch increments, applied to data from experiment 1, subjects 1 (a) and 2 (b).

3.5.3 Experiment 5: Impedance signal changes with RLC circuit

The RLC circuit with the largest bandwidth, configuration 3 ($L_c = 44$ mH and $C_c = 1.45$ nF), provided the highest tolerance when manually tuning the frequency of the drive current using the benchtop waveform generator due to the broadness of the peak in its frequency response. However, the relatively low capacitance of the RLC capacitance component, $C_c$, meant the resonant frequency was highly susceptible to variations in the nerve capacitance. Conversely, in configuration 1 ($L_c = 13$ mH and $C_c = 6.25$ nF) the relatively narrow peak and high capacitance, respectively, provided the lowest tolerance when manually tuning the frequency of the drive current and lowest susceptibility to variation in the nerve capacitance. Configuration 2 ($L_c = 22$ mH and $C_c = 3.45$ nF) provided a balance between tuning tolerance and susceptibility to variation in nerve capacitance, and, therefore, was used in the RLC experiment.

In subjects 6 – 9 the manual jumper settings which set the RLC capacitance component, $C_c$, were 4.14, 4.1, 4.14, and 4.1 nF, respectively. In subjects 6 – 9, the RLC experiment was conducted at 16.9002, 17.1002, 17.3002, and 17.3002 kHz, respectively, after manual tuning of the waveform generator to find the resonant peak. In subjects 6 and 7 the RLC circuit was on for RLC set 2, while in subjects 8 and 9 the RLC circuit was on for RLC set 1, Fig 4d. The impedance signal amplitude peak (+/-3σ noise) with the RLC circuit off and on, respectively, were -0.034 and -0.049 % (+/-0.007 %), -0.016 and -0.021 % (+/-0.004 %), -0.051 and -0.070 % (+/-0.006 %), and -0.033 and -0.039 % (+/-0.005 %), Figs 13a - 13e. The impedance signal peak amplitude was, therefore, consistently larger with the RLC on, with improvement in the peak amplitude (+/-3σ noise), with respect to the RLC off condition, of 44 % (+/-15 %), 34 % (+/-30 %), 37 % (+/-8.6 %), and 16 % (+/-19 %), Table 2.

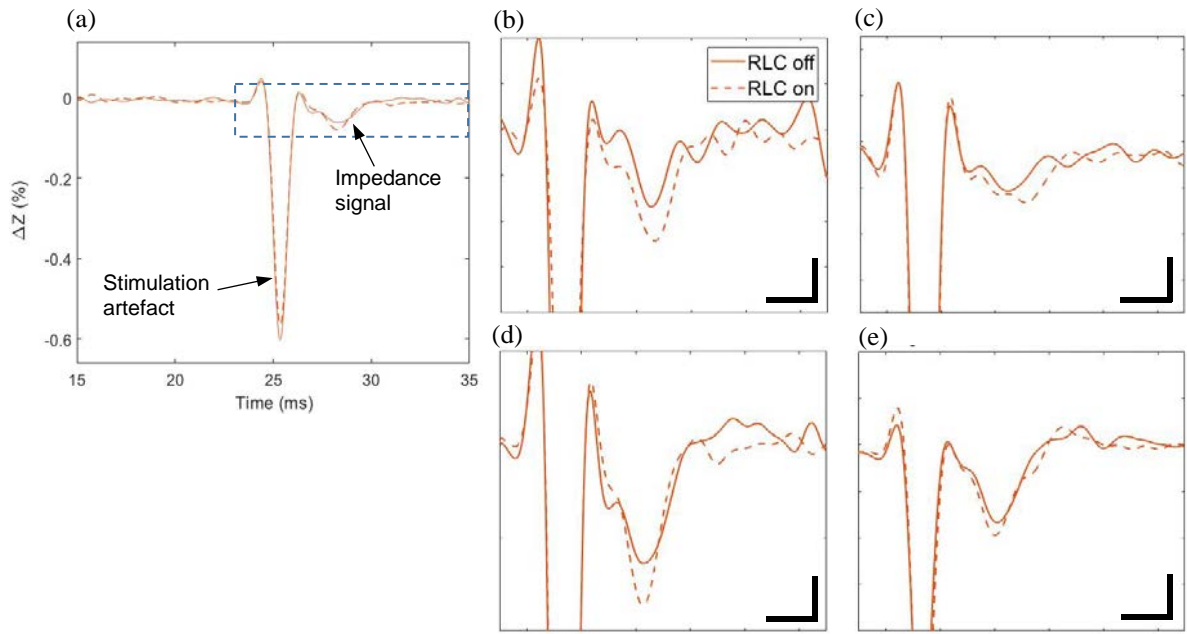

Figure 13: Impedance signal and stimulation artefact in subject 8 (a), with a dashed box showing the 23 to 35 ms and -0.08 to +0.04 % range in (b - e). Impedance signals in subjects 6 – 9 (b - e), showing differences with RLC on and RLC off across the drive current source, where the vertical and horizontal black bars are 0.02 % change in impedance and 2 ms in time, respectively. The RLC circuit was on in RLC set 2 in subjects 6 and 7, (b, c), and in RLC set 1 in subjects 8 and 9 (d, e). An increase in impedance signal amplitude is visible in all experiments.

Table 2: Summary of impedance signal amplitude results of experiment 5 using the RLC circuit. Data collected in RLC set 2 (*) is expected to exhibit a 10 to 20 % increase due to the sampling method.

| Experiment | Impedance signal (%) | | | Relative improvement |
| --- | --- | --- | --- | --- |
| | RLC off | RLC on | +/- 3σ noise | +/- 3σ noise (%) |
| 6 | -0.034 | -0.049* | +/- 0.007 | 44 +/- 15 |
| 7 | -0.016 | -0.021* | +/-0.004 | 33 +/- 30 |
| 8 | -0.051* | -0.070 | +/-0.006 | 37 +/- 8.6 |
| 9 | -0.033* | -0.039 | +/-0.005 | 16 +/- 19 |

## 4. Discussion

In experiment 1, the 6σ system noise before averaging, of 1 mV (170 µV 1σ), is considerable in comparison to the size of the impedance signal amplitude, of 5 to 100 µV. Ensemble averaging of several hundred epochs was an effective means to generating adequate SNR for this investigation, Fig 7, but the use of lower noise hardware to increase the SNR would improve confidence in the results. Lower noise systems, such as that used in [14], required ensemble averaging of 840 epochs to produce an SNR of 8, which is adequate for EIT reconstruction of neural activity in peripheral nerve [14].

In experiment 2, temporal changes to the impedance signal amplitude and delay between stimulation and impedance signal, observed in subjects 1 – 2, were significant, Fig 8c and 8d. The reasons for the amplitude and delay effects are thought to be the lack of natural perfusion and cooling of the tissue. For the latter, modelling in [4] indicates that reducing the temperature reduces action potential propagation velocity in unmyelinated fibres, which might also extend to myelinated fibres. Good agreement between results of the temporal change experiments provided confidence in the polynomial fit, equation 5, and its subsequent use to correct frequency response data in experiment 3. The duration of the frequency response experiment was limited in length to 1200 epochs, or 600 s, to avoid the largest impedance signal reduction observed between 600 and 900 s, and so offset the reduced SNR which was expected from ensemble averaging of only 200 epochs at each frequency. The delay between stimulation and impedance signal elongates and reduces the amplitude of the impedance signal during ensemble averaging and, thereby, reduces the frequency components of the extracted impedance signal. This reduction in frequency components was not evaluated in the current study, with band pass filter bandwidth of 1500 Hz [14] used across all experiments during signal processing.

In experiment 3, the frequency response of subjects 3 – 5 showed good agreement, Fig 9, with an increase in impedance signal in the frequency range 6 – 8 kHz, and decline from 8 – 18 kHz. Modelling of myelinated fibres in [5] produced an increase in impedance signal between 6 – 8 kHz and decline either side of this range when using phase-antiphase subtraction to extract the impedance signal [22], and a local maximum at 12 kHz when using the band-pass filter plus Hilbert transform method employed in the current study. The frequency response of peripheral nerve *in-vivo* under a transverse current in [10] showed an increase in impedance signal between 6 – 10 kHz. The similar frequency response roll off in both transverse and longitudinal drive currents indicates the phsyiological basis is independent of the drive current orientation, an observation also made in [23] on dorsal column of cat. The significant decrease in signal amplitude in the frequency range 17 +/- 1 kHz was considered a good candidate for RLC experiments due to the expected presence of significant capacitive charge transfer across the myelin sheath. Frequency response experiments on brain tissue [15, 16] showed the impedance signal peaks and then declines before plateuing at approximately 7 to 20 % of its peak value, even at frequencies a factor of 6 higher than that of the peak impedance signal. This plateuing indicates that frequencies significantly above 18 kHz may also be good candidates for RLC experiments in peripheral nerve.

In experiment 4, for a 0.6 mm radius by 3 mm length nerve - based on the nerve cuff radius and electrode spacing, respectively - the estimated nerve capacitance, of 0.75 to 0.8 nF, corresponds to a specific capacitance of 2 to 2.12 µF/m, whereas estimated nerve resistance, of 7.9 to 8.7 kΩ, correspdonds to a specific resistance of 2.98 to 3.28 Ω.m. The specific capacitance is a factor of approximately 5 lower than that derived from modelling in [20], whereas the specific resistance shows good agreement. The frequency response roll off of the nerve in nerve cuff predicted in modelling, at frequencies above 4 kHz, Fig 11b, generally agreed with frequency response data in experiment 3 for subjects 3 – 5, with a roll-off at frequencies above 6 kHz, Fig 9. However, the steepness of the frequency response roll-off in modelling, with -6 dB reduction in impedance signal at 40 kHz, did not agree with data from experiment 3, subjects 3 – 5, where the -6 dB reduction was located at approximately 14 kHz. This disagreement suggests either degredation in tissue health between explantation and measurement has affected the values, or the simplified electrical model of nerve tissue, Fig 5a, does not accurately model impedance changes in neural tissue during neural activity, or a combination of both.

In experiment 5, the range of impedance signal amplitudes across subjects 6 – 9 with the RLC off, of -0.014 to -0.051 %, Fig 13, are broadly in agreement with those obtained in frequency response experiment 3, of -0.014 to -0.030, Fig 9 (within which data from subject 3 has been divided by 2). The variation in impedance signal amplitudes between subjects is due to differences in the tissue morphology and, as a result, differences in the proximity of the recording electrodes to underlying neural activity, and is consistent with results in [14]. The large and consistent relative improvements in the impedance signal amplitude across subjects 6 – 9 in the RLC experiment, of 44, 33, 37, and 16 %, Table 2, provide confidence in the RLC circuit design and implementation despite relatively high levels of uncertainty introduced by the noise (+/- 15, 30, 8.6, and 19 %, respectively) and sampling method (10 – 20 % increase in set 2). In addition to uncertainties from noise and sampling method, a further reason for the broad range of values observed in the relative improvement, of 16 to 44 %, is the frequency tuning resolution of 100 Hz and accuracy of visually identifying the peak, with the latter estimated at 300 Hz.

Application of this +/- 350 Hz range to the modelled frequency response of RLC circuit configuration 2, Fig 11b, produces a reduction in the relative improvement in the range 1 to 4 %. The flatter peak of RLC circuit configuration 3 would mitigate this tuning error. The relative improvements in impedance signal amplitude in experiments were predominantly higher than the 17 % predicted in modelling, Fig 11b, which may be explained by the poor ability of the simplified electrical model of nerve tissue in modelling impedance change in neural tissue.

At 17 kHz, the polynomial fit to the frequency response in experiment 3 indicated an approximately -8.5 dB reduction in impedance signal amplitude from the maximum at 6 – 8 kHz. The relative improvements in impedance signal amplitude produced using the RLC circuit only partially ameliorates this reduction. Repetition of the RLC component selection process, Fig 11a and 11b, for bandwidths of 3.6 kHz and resonant ferquencies 30 kHz ($L_c$ = 7 mH and $C_c$ = 3.5 nF) and 50 kHz ($L_c$ = 2.5 mH and $C_c$ = 3.5 nF), Fig 14, reinforces the observation that the maximum impedance signal is not obtainable using the RLC circuit. This limitation is due to the presence of the passive nerve resistance, $R_p$, which is unaffected at all frequencies regardless of capacitive charge transfer, and the electrode resistive element, $R_e$, which is in parallel with a compararatively high capacitance and so dominated by capacitive charge transfer at lower frequencies. In reality, the series resitive element $R_p$ contains contribution from all series resistances in the current source path, not just the passive resistance of the neural tissue. Despite this partial amelioration in the recoverable impedance signal, a relative improvement to the impedance signal amplitude in the range 16 to 44 % provides a meaningful increase to the SNR, resulting in an extension to the range of frequencies which can be implemented in a FDM EIT system, and reducing the number epochs required to achieve an adequate SNR for EIT imaging.

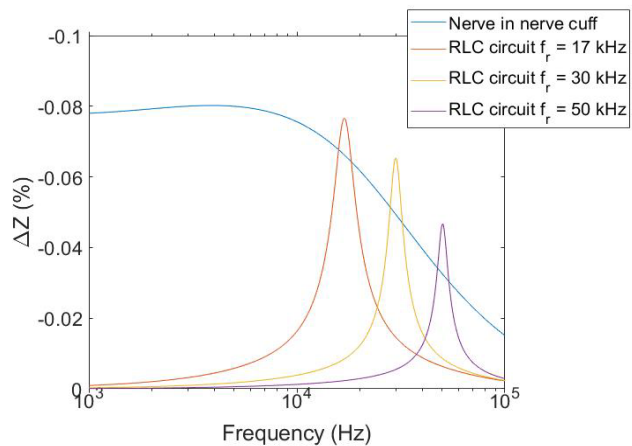

Figure 14: The frequency response of the impedance signal generated from a 0.1 % change in the nerve fibre resistance in equivalent electrical circuits of the nerve and nerve cuff circuit and the RLC circuits with bandwidths of approximately 3.6 kHz and resonant ferquencies 17, 30 and 50 kHz. The peak impedance signal at 6 – 8 kHz not fully ameliorated by the RLC circuits.

While the RLC circuit configuration presented here is a novel means of ameliorating capacitive charge transfer across the myelin sheath, studies using charge balanced sinusoidal signals for excitation of neural tissue produced strength-threshold curves which, approximately, show an inverse relationship to the frequency response roll-off in neural tissue [24]. This inverse relationship indicates that capacitive charge transfer across myelin or neuron membranes does not contribute significantly to excitation of neural tissue, and, consequently, the EIT drive current amplitude may be increased at higher frequencies to improve the SNR without risking excitation of the tissue under study. However, achieving higher SNR by increasing drive current amplitude, particularly with multiple concurrently active drive currents as in FDM EIT, must be carefully considered as an option as it increases the overall current density in passive tissues and satellite cells, and increases tissue heating near the drive electrodes.

Two final observations from modelling of the RLC circuit are, firstly, the consistent phase response with RLC circuits, of 180º at the resonant peak frequency, due to the mutual cancellation of the inductive and capacitive elements. In phase division multiplexing (PDM), the phase contribution of each element in the system must be characterised and included in the signal processing arithmetic [25]. Therefore, control of the phase across the drive electrodes and neural tissue, by using the RLC implementation presented here, would be of benefit; though with the caveat that phase contributions from the measurement electrodes must still be contended with. Secondly, the frequency response of the RLC circuit acts as a bandpass filter around the drive current frequency, thereby removing much of the wideband noise which would otherwise be present in the neural activity, typically extracted from peripheral nerve data using a 1 to 2 kHz (-3 dB) low pass filter [10, 14], and in the band pass filter range of other drive currents in an FDM EIT system [14].

*4.1 Limitations*

The current study is intended to act as a proof of concept of a parellel RLC circuit implemented across an EIT current source, and description of the design process. As such, RLC circuit design and experiments were limited to a single frequency. Further investigation to determine the relative improvement provided by the RLC circuit across a frequency range, interactions between the neural tissue and RLC circuit, and implementation of RLC circuit design process across more realistic neural tissue models, would all further understanding of this novel design. It is also noted that manual tuning of the waveform generator to find the resonant frequency during RLC experiments could be improved using hardware capable of sweeping a frequency range with high frequency resolution and automatically locating the local maximum.

**5. Conclusion**

The frequency response of peripheral nerve was characterised under a longitudinal current in the frequency range 4 – 18 kHz, with data corrected for temporal changes in the impedance signal during *in-vitro* experiments on sciatic nerve of rat. A novel current source implementation was developed and tested which implements a parallel RLC circuit across the nerve and nerve cuff by introducing capacitance and inductance components. The RLC circuit implementation produces a relative improvement in the impedance signal, observed during neural activity, of 16 to 44 % by ameliorating the effects of capacitive charge transfer within the neural tissue. The resultant increase in impedance signal at high frequencies improves spatial resolution in EIT imaging by increasing the number of parallel drive currents which can be implemented in an FDM EIT system, and aids the long term goal of a real-time FDM EIT system by reducing the number of epochs in ensemble averaging.

**Acknowledgements**

The authors would like to thank David Holder, Kirill Aristovich and Adam Fitchett, from the EIT Research Group at University College London, for providing essential hardware in support of this research and for help interpreting results; Darren Svirskis and Mahima Bansal from the School of Pharmacy, The University of Auckland for their help with PEDOT coating electrodes; and staff at the School of Biological Sciences, The University of Auckland for their help with animal handling